\documentclass[preprint,showpacs,preprintnumbers,nofootinbib,amsmath,amssymb]{revtex4-1}
\usepackage{graphics,epsfig,subfigure}
\usepackage{epstopdf}
\usepackage{color}
\usepackage{mathrsfs}

\begin{document}
\renewcommand{\baselinestretch}{1.3}
\newcommand\be{\begin{equation}}
\newcommand\ee{\end{equation}}
\newcommand\ba{\begin{eqnarray}}
\newcommand\ea{\end{eqnarray}}
\newcommand\nn{\nonumber}
\newcommand\fc{\frac}
\newcommand\lt{\left}
\newcommand\rt{\right}
\newcommand\pt{\partial}
\newcommand\tc{\textcolor[rgb]{1,0,0}}

\title{Gravitational waves and extra dimensions: a short review}

\author{Hao Yu\footnote{yuh13@lzu.edu.cn}, Zi-Chao Lin\footnote{linzch12@lzu.edu.cn}, Yu-Xiao Liu\footnote{liuyx@lzu.edu.cn, corresponding author}}

\affiliation{
Institute of Theoretical Physics $\&$ Research Center of Gravitation, Lanzhou University, Lanzhou 730000, China}

\begin{abstract}
{We give a brief review on the recent development of gravitational waves in extra-dimensional theories of gravity. Studying extra-dimensional theories with gravitational waves provides a new way to constrain extra dimensions. After a flash look at the history of gravitational waves and a brief introduction to several major extra-dimensional theories, we focus on the sources and spectra of gravitational waves in extra-dimensional theories. It is shown that one can impose limits on the size of extra dimensions and the curvature of the universe by researching the propagations of gravitational waves and the corresponding electromagnetic waves. Since gravitational waves can propagate throughout the bulk, how the amplitude of gravitational waves decreases determines the number of extra dimensions for some models. In addition, we also briefly present some other characteristics of gravitational waves in extra-dimensional theories.}
\end{abstract}

\maketitle

\section{Introduction}

On 11 February 2016, the LIGO and Virgo Scientific Collaborations announced that they detected, directly, a transient gravitational wave (GW) signal on 14 September 2015, which was named as GW150914~\cite{Abbott:2016blz}. The explosive news quickly caught the attention of the scientific community. Based on the data of GW150914 and the several subsequent GW events~\cite{Abbott:2016nmj,Abbott:2017vtc,Abbott:2017gyy,Abbott:2017oio,
TheLIGOScientific:2017qsa}, many related studies have been rapidly developed.

As we know, the current accuracy of observation of GWs is not enough to constrain modified gravity theories if we do not consider the combination with their electromagnetic counterparts. Therefore, simultaneous detection of GWs and their counterparts is particularly significant. Although the Gamma-Ray Burst Monitor (GBM) detected a weak gamma-ray burst in the GW150914 event, most scientists believe that the electromagnetic signal does not come from the source of GW150914 because the source location of GW150914 could only be confined to an arc on the sky. Two years later, the LIGO and Virgo Scientific Collaborations and the GBM instrument detected the GW170817/GRB170817A event~\cite{TheLIGOScientific:2017qsa}, which undoubtedly dispels people's doubts about the simultaneous detection of GWs and electromagnetic signals. The detection of the GW170817/GRB170817A event marks the arrival of multi-messenger astronomy~\cite{GBM:2017lvd}, and brings the chance to most modified gravity theories.

As one of the modified gravity theories, Horndeski gravity~\cite{Horndeski:1974wa} has been widely studied in recent years. Since GW150914 was detected, the study of GWs in Horndeski gravity has involved many aspects: propagations of GWs~\cite{Arai:2017hxj}, polarizations of GWs~\cite{Hou:2017bqj,Gong:2018ybk,Hou:2018mey}, primordial GWs~\cite{Nunes:2018zot}, and so on. As for $f(R)$ gravity~\cite{Vainio:2016qas,Moradpour:2017lcq,Sharif:2017ahw,Lee:2017dox,
Jana:2018djs} and scalar-tensor theory~\cite{Scharre:2001hn,Sotani:2004rq,Yunes:2011aa,Bettoni:2016mij,
Shao:2017gwu,Gong:2017bru,Sakstein:2017xjx,Gong:2017kim,Bartolo:2017ibw}, the properties and applications of GWs in various models are also hot topics. In addition, the detection of GWs can give a limitation on the mass of gravitons in massive gravity, especially the upper limit for the mass of gravitons. This is mainly because the presence of the mass term could influence the spectrum and speed of GWs~\cite{Gumrukcuoglu:2012wt,deRham:2016nuf,Lin:2017fec,Fujita:2018ehq,
Perkins:2018tir}.

As a new tool to explore the universe, GWs can almost involve every aspect of cosmology. We can use GWs to detect dark matter, especially primordial-black-hole dark matter~\cite{Chen:2016pud,Dev:2016hxv,Clesse:2016ajp,Wang:2016ana,
Garcia-Bellido:2017fdg,Kovetz:2017rvv,
Gong:2017qlj,Cai:2017buj,Jung:2017flg,Flauger:2017ged,Chen:2018czv}. GWs as standard sirens are closely related to dark energy~\cite{Caprini:2016qxs,Caldwell:2018feo,BeltranJimenez:2018ymu,
Ezquiaga:2018btd,Creminelli:2018xsv,Du:2018tia}. With these GW events, we can either impose constraints on dark energy models~\cite{Caprini:2016qxs,Du:2018tia}, or improve the constraints on the propagation speed of GWs~\cite{BeltranJimenez:2018ymu}. Furthermore, we can also get more properties of inflation
\cite{Liddle:1992wi,Baumann:2006cd,Dufaux:2007pt,Cook:2011hg,Senatore:2011sp,
Liu:2017hua},
gravitational lenses~\cite{
Frieman:1994pe,Takahashi:2005sxa,Collett:2016dey,Fan:2016swi,Hannuksela:2019kle}, and
phase transitions~\cite{Kosowsky:1992rz,Grojean:2006bp,Caprini:2009yp,Caprini:2015zlo,
Ellis:2018mja,Cai:2018teh} through GW observations.

For extra-dimensional theories, people have always focused on detecting extra dimensions with high-energy experiments. But now, the detection of GWs provides a new way to detect extra dimensions. Combining the two methods of detecting extra dimensions, one can obtain more strict constraints on extra-dimensional theories. Here we only discuss how to use the latter to detect extra dimensions. First, in some extra-dimensional theories, the number of extra dimensions could affect the amplitude attenuation of GWs, which is widely recognized in extant literatures~\cite{Frolov:2002qm,Kobayashi:2003cn,BouhmadiLopez:2004ax,Ghayour:2012nf,
Hiramatsu:2004aa,Kobayashi:2005dd,Clarkson:2006pq}. Second, the size of extra dimensions could affect the size of the shortcut that a gravitational signal takes in the bulk~\cite{Caldwell:2001ja,Abdalla:2001he,Abdalla:2002ir,Abdalla:2002je,
Abdalla:2005wr}. These two features of extra dimensions are vital in the process of detecting extra dimensions through GWs. Although the current detection of GWs is not accurate enough, some constrains on the parameters of extra-dimensional models based on the existing data have been obtained~\cite{Yu:2016tar,Visinelli:2017bny,Pardo:2018ipy,Qiang:2017luz}. On the other hand,
as early as a few decades ago, it has been thought that these features of GWs in extra-dimensional theories can be used to solve cosmological problems (the most prominent one is to explain the horizon problem with shortcuts through the bulk~\cite{Chung:1999xg,Chung:1999zs,Ishihara:2000nf}).

The structure of the short review is as follows. In the second part, we briefly introduce some major events in the research and detection of GWs and several important extra-dimensional theories. Next, in Section III we introduce the sources of GWs and the characteristics of the corresponding spectra. We mainly focus on the difference between the spectra of GWs in extra-dimensional theories and the spectrum in standard general relativity. In Section IV we turn our attention to the shortcuts of GWs in extra-dimensional theories. In Section V we discuss how to use GWs to detect the size and number of extra dimensions. Section VI is dedicated to some other GWs. Our summary and outlook are given in Section VII.

\section{Background}
\label{sec1}

\subsection{History of Gravitational Waves}
\label{sec11}
The concept of GWs was first proposed by Oliver Heaviside in 1893 based on the analogy that gravity and electricity all satisfy the inverse-square law, and he also found that the produced GWs travel at a finite speed. Later, in 1905, Henri Poincar\'{e} pointed out that GWs should propagate at the speed of light.

In 1915, Einstein published general relativity (GR). The next year, he predicted the existence of GWs, deduced the wave equation satisfied by GWs in general relativity, and found that the speed of GWs is indeed the speed of light (now we know that there were some errors in Einstein's deduction at that time, and he arrived at the correct formula for gravitational radiation until 1918). However, his work was questioned by some scholars and he also had no confidence in his own results (see Ref.~\cite{Cervantes-Cota:2016zjc}). In 1936, Einstein revisited the topic of GWs with his assistant Nathan Rosen and submitted a paper to Physical Review claiming that there exists no real GWs at all because all the solutions of Einstein's equations would have singularities. This time, Einstein made another mistake of using bad coordinates, which was corrected soon by Howard P. Robertson~\cite{Cervantes-Cota:2016zjc,Kennefick:2007zz}. In a sense, Einstein's suspicions about the truth of GWs promoted this field to move forward. In the second year after Einstein's death, people made a major breakthrough in experimental observation of GWs. In 1956, Felix Pirani re-described GWs with a manifestly observable Riemann curvature tensor, which remedied the confusion caused by the use of various coordinate systems. He also proved that GWs are detectable since they could change the proper distance between at least two free-falling test particles (the test particles should have very low masses and their own gravity can be ignored). In the next year, Richard Feynman solved the problem of whether GWs could transmit energy during the first ``GR'' conference. Since then the research on GWs entered the era of detection~\cite{Kennefick:2007zz,Cervantes-Cota:2016zjc}.

Inspired by the work of Felix Pirani, Joseph Weber of the University of Maryland designed and set up the first GW detector, known as Weber bars. In 1969, Weber claimed that the first GW signal was detected, but it was soon denied by himself and other (theoretical and experimental) physicists. Although he did not detect any GW signal with his device, his concept of using a rod-like detector to detect GWs was later widely accepted and improved.

As astronomers discovered quasars in the late 1950s and pulsars in 1967, the hope of detecting GWs was pinned on quasars and pulsars. These celestial bodies belong to neutron stars or black holes, which are very massive compact objects. We must consider GR when describing their gravitational properties. In 1974, Russell Alan Hulse and Joseph Hooton Taylor Jr. discovered the first pulse binary named Hulse-Taylor pulsar (or PSR B1913+16). Their observations in subsequent years showed that the orbital period of the binary system was decaying gradually and they were getting closer to each other. These phenomena could be explained by the gravitational radiation predicted by GR~\cite{Taylor:1979zz,Taylor:1982zz}. Therefore, the study of PSR B1913+16 is the first evidence that indirectly proves the existence of GWs.

During this period, the experiment to directly detect GWs had also advanced to a new stage: using a laser interferometer to detect GWs. This method was first proposed by the Russian physicists Mikhail Evgen'evich Gertsenshtein and  Vladimir Ivanovich Pustovoit in 1962. And the first prototype was built in the 1970s by Robert L. Forward and Rainer Weiss. After 150 hours of observation, Forward reported that no GWs were observed.

In 1984, the California Institute of Technology and the Massachusetts Institute of Technology signed a contract agreeing to cooperate in the design and construction of the Laser Interferometer Gravitational-Wave Observatory (LIGO). In 1990, the LIGO program was approved to build the same detector in Livingston and Hanford, respectively, in order to remove unrelated signals. In 2002, LIGO began to detect GWs for the first time and in 2010, it ended collecting data. During this period, no GWs were detected, but they gained a lot of valuable experience. Between 2010 and 2014, LIGO was redesigned and rebuilt to improve sensitivity by more than 10 times. After the upgrade, it was renamed ``Advanced LIGO'' (aLIGO) and was restarted in 2015. Another large interferometer Virgo (which was built in 1996 in Italy) was also completed in June 2003 and several data collections were conducted between 2007 and 2011. Since 2007, Virgo and LIGO signed a cooperation agreement to jointly process detector data and publish detection results.

After years of unremitting efforts, on 11 February 2016, the LIGO and Virgo teams announced that GWs were detected for the first time on 14 September 2015~\cite{Abbott:2016blz}. This event (namely, GW150914) originated from a pair of merging black holes $410^{+160}_{-180}$ Mpc away from the Earth. By the end of 2017, LIGO and Virgo had detected several GW events~\cite{Abbott:2016nmj,Abbott:2017vtc,Abbott:2017gyy,Abbott:2017oio,
TheLIGOScientific:2017qsa}. It is worth mentioning that, in these GW events, GW170817 is the first time that LIGO and Virgo detected a GW generated by the merger of two neutron stars. Just 1.7 seconds later, a short gamma ray burst (GRB170817A) was discovered by the GBM and it is likely that these two signals come from the same source. The detection of GW170817 and its electromagnetic counterpart is the first direct evidence that supports the link between mergers of binary neutron stars and short gamma ray bursts.

\subsection{Several Extra-dimensional theories}
\label{sec12}

In order to unify electromagnetism and gravity, Gunnar Nordstr\"{o}m first proposed the conception of extra dimensions in 1914~\cite{Gunnar1,Gunnar2}. Then Theodor Kaluza and Oskar Klein introduced a five-dimensional space-time theory, dubbed Kaluza-Klein (KK) theory in the 1920's~\cite{Klein1,Klein2,Klein3}. In this well-known theory, the extra dimension is assumed as a compact circle. A consequence of this assumption is that every quantity defined by this compact extra dimension obeys a periodic boundary condition. The metric tensor could be Fourier expended to a series of KK modes, and the effect of the non-zero KK modes would vanish due to the periodic boundary condition. The most prominent feature of the KK theory is that it could recover both the electromagnetism and GR in four-dimensional space-time. Since this theory is a pure theory of gravity, the four-dimensional electromagnetism could be regarded as a pure gravitational effect.

Although the KK theory is very successful in unifying gravity and electromagnetism, it has some problems when we consider a coupling between gravity and matter fields. To make this clear, we first study a massless scalar field $\phi(x^{\mu},y)$ in the five-dimensional space-time. Since it is a five-dimensional field existing in the bulk, its five-dimensional Klein-Gordon equation is given by
\begin{eqnarray}
\square^{(5)}\phi\equiv(\partial_{\mu}\partial^{\mu}+\partial_{y}^{2} )\phi=0.
\end{eqnarray}
With the periodic boundary condition on the extra dimension $y$, the solution of this scalar field is
\begin{eqnarray}
\phi=\sum_{n=0}^{\infty}e^{ip_{\mu}x^{\mu}}e^{iny/R_{\text{ED}}},
\end{eqnarray}
where $R_{\text{ED}}$ is the radius of the extra dimension, the integer $n=0,\pm1,\pm2,\ldots$ denotes the mode of the scalar field, and the angular momentum $p_{\mu}$ obeys
\begin{eqnarray}
p_{\mu}p^{\mu}=-\frac{n^{2}}{R_{\text{ED}}^{2}}.
\end{eqnarray}
In this case, the massless bulk scalar is the combination of the zero mode and a series of massive KK modes (the mass spectrum satisfies $m_{n}=|n|/R_{\text{ED}}$). From the point of view of different local observers located along the extra dimension, the bulk scalar could obtain charge through the gauge translation on the fifth coordinate. This mechanism provides a natural way to introduce charge quantisation in the KK theory~\cite{Klein4}. The consistency between the charge element obtained from this mechanism and the electromagnetic coupling constant detected in experiments requires the radius of the extra dimension to be $10^{-33}\,\text{m}$, which is closed to the Planck length $\ell_{\text{Pl}}~(\sim10^{-35}\,\text{m})$.

However, there is no reason to impose the mass of the bulk field to be zero. For a massive bulk scalar field, the mass spectrum should be modified as
\begin{eqnarray}
m_{n}=\sqrt{M_{0}^{2}+\frac{n^{2}}{R_{\text{ED}}^{2}}},
\end{eqnarray}
where $M_{0}$ is the mass of the bulk scalar field. It is found that, due to the length scale of $R_{\text{ED}}$, the masses of the non-zero KK modes are far beyond the capacity of particle collision experiments. In other words, the non-zero KK modes of the bulk matter field are impossible to be found in experiments (the zero mode corresponds to the four-dimensional elementary particle). The parameter $M_{0}$ should be fixed at the electroweak scale~\cite{Raychaudhuri1}. As shown in Refs.~\cite{Raychaudhuri1,Liu1}, the zero mode could not be charged through the mechanism referred above, which means that all the four-dimensional elementary particles are neutral. Obviously, it is contradict to the reality. On the other hand, the requirement that the charge element must be much smaller than the electromagnetic coupling constant could efficiently suppress the magnitude of the masses of KK modes, but it also makes the electromagnetic field obtained by dimensional reduction be weakly coupled to the ordinary matter field (it is no longer the electromagnetic field we observe in four-dimensional space-time), which is totally deviated from the fundamental starting point of the KK theory.

In 1983, a well-known extra-dimensional theory called domain wall theory was proposed by Valerii A. Rubakov and Mikhail E. Shaposhnikov~\cite{Rubakov1,Rubakov2}. In this theory, an infinite extra dimension together with a bulk scalar field are introduced. The authors found that an effective potential well along the extra dimension could localize the energy density of the scalar field in the well. Therefore, the energy density of the scalar field constructs a three-dimensional hypersurface, dubbed domain wall, embedded in the five-dimensional space-time. Unlike the KK theory, the particles in the Standard Model are arisen from the reduction of the perturbation of the bulk scalar field. Moreover, the KK modes of the bulk fermion field could be constructed by the shape of the scalar potential instead of introducing a compact extra dimension. In this case, the zero mode and lower massive KK modes of the bulk fermion field could be trapped on the domain wall. The Standard Model particles could only travel on the domain wall at a speed less than or equal to the speed of light. On the other hand, the KK modes of the fermion field with mass square higher than the potential well could move along the fifth dimension freely. In phenomenology, observers on the domain wall could observe that these KK modes escape from the domain wall, resulting in a leakage of energy during particle collisions. Nevertheless, since the fifth dimension is infinite and the five-dimensional space-time is flat, it is hard to localize the zero mode of gravity on the domain wall. The gravitational force between two particles will deviate from the inverse square law, being proportional to $1/r^{D-2}$ with $D$ the dimensionality of the space-time.

Now, before referring to the next extra-dimensional model, we remind a long-standing puzzle in particle physics, which is called as the hierarchy problem. To explain the issue, we recall the Einstein's equations in four-dimensional space-time as follows~\cite{Raychaudhuri1}
\begin{eqnarray}\label{E1}
\frac{1}{\ell_{\text{Pl}}}\Big(R_{\mu\nu}-\frac{1}{2}g_{\mu\nu}R_{(4)}\Big)
=8\pi\ell_{\text{Pl}}T_{\mu\nu},
\end{eqnarray}
where we have used $G=\ell_{\text{Pl}}^{2}$. It is obvious that, if the relativistic energy of the matter field is low, it is a good approximation to regard the space-time as a flat one and neglect the right hand side of Eq.~\eqref{E1}. However, if the energy of the matter field is relatively high, i.e., $\ell_{\text{Pl}}E\sim1$, the curvature of the space-time cannot be ignored. It means that, when the energy scale of the matter field reaches the Planck scale, both the gravitational and electroweak interactions should be counted in the quantum field theory defined in the flat space-time. In this case, a serious issue arises: it seems hard to explain the huge discrepancy between the Planck scale $M_{\text{Pl}}\sim10^{19}\,\text{GeV}$ and the electroweak scale $M_{\text{EW}}\approx246\,\text{GeV}$ in the Standard Model.

In a higher-dimensional theory with compact extra dimensions, the Einstein's equations are~\cite{Raychaudhuri1}
\begin{eqnarray}
\frac{1}{\ell^{1+d}_{\ast}}\Big(R_{MN}-\frac{1}{2}g_{MN}R_{(4+d)}\Big)=8\pi
\ell_{\ast}T_{MN},
\end{eqnarray}
where $d$ is the number of the compact extra dimensions, and $\ell_{\ast}$ is the bulk Planck length. Assuming that all the compact extra dimensions have the same size, $\ell_{\ast}$ is therefore related to Planck mass through the relation~\cite{Liu1}
\begin{eqnarray}\label{MM}
M^{2}_{\text{Pl}}=M_{\ast}^{2+d}(2\pi R_{\text{ED}})^{d},
\end{eqnarray}
where $M_{\ast}=\ell_{\ast}^{-1}$ is the bulk Planck mass and $R_{\text{ED}}$ is the radius of the compact dimensions. From gravity and particle physics experiments (see Refs.~\cite{YangLuo:2012PRL,TanLuo:2016PRL} and Fig.~6.1 in Ref.~\cite{Raychaudhuri1}), the radius of the compact extra dimension and the bulk Planck mass should be constrained to $R_{\text{ED}}<60\,\mu\text{m}$ and $M_{\ast}>1\,\text{TeV}$, respectively. Therefore, the hierarchy problem could be solved by tuning the scale of $M_{\ast}$ to electroweak scale through $R_{\text{ED}}$ and $d$. If the model has less than six extra dimensions and the bulk Planck mass ranges from $1\,\text{TeV}$ to $10\,\text{TeV}$, then the radius of the compact extra dimensions should at least be the same as the size of a neutron, i.e., $R_{\text{ED}}\sim10^{-15}\,\text{m}$. Therefore, the particles produced from high-energy particle collision experiments could access the extra dimensions. Until now, people have not found such signal in any high-energy experiment, which might indicate that the radius of the extra dimensions is less than $10^{-18}\,\text{m}$. So, we will finally obtain a theory either with incredibly large numbers of extra dimensions or without capacity of curing the hierarchy problem.

At the end of last century, Nima Arkani-Hamed, Savas Dimopoulos, and Georgi R. Dvali (ADD) realized that, if the particles in the Standard Model are confined on a three-dimensional hypersurface by some unknown mechanism, the leakage of energy will never occur in particle experiments and the contradiction above will naturally vanish~\cite{ArkaniHamed1}. Based on this inspiration, they successfully constructed relatively large compact extra dimensions in their well-known ADD model. But, is ADD model really safe from the hierarchy problem? The answer is no. Recalling Eq.~\eqref{MM}, the ratio between the radius of the extra dimensions and the basic unit of the extra-dimensional model (i.e., the bulk Planck length $\ell_{\ast}$) is given by~\cite{Liu1}
\begin{eqnarray}
\frac{R_{\text{ED}}}{\ell_{\ast}}=\frac{1}{2\pi}
\Big(\frac{M_{\text{Pl}}}{M_{\ast}}\Big)^{\frac2d}\gtrsim10^{\frac{26}{d}-1}.
\end{eqnarray}
Apparently, there still exists a large hierarchy between the two fundamental quantities. Therefore, ADD model just transforms the hierarchy problem into a new insight.

We now know that, for a well-defined extra-dimensional theory, it should be able to, on one hand, explain the large discrepancy between the Planck scale and the electroweak scale and, on the other hand, not bring new hierarchy between the radius of extra dimensions and the fundamental length scale. The breakthrough came out from the work of Lisa Randall and Raman Sundrum (RS)~\cite{Randall1}. In their well-known RS-I model, they introduced a warped structure to the compact extra dimension. Then there appears a warp factor $A(y)$ in the five-dimensional metric~\cite{Randall1}:
\begin{eqnarray}
ds^{2}=e^{2A(y)}\eta_{\mu\nu}dx^{\mu}dx^{\nu}+dy^{2}.
\end{eqnarray}
With this metric, Eq.~\eqref{MM} needs to be rewritten as
\begin{eqnarray}
M^{2}_{\text{Pl}}=\frac{M_{\ast}^{3}}{k}(1-e^{-2k\pi R_{\text{ED}}}),
\end{eqnarray}
where $k$ is a parameter with the dimension of mass. Moreover, the compact warped structure of the extra dimension naturally leads to two special points, i.e., $y=0$ (the position of Planck brane) and $y=\pi R_{\text{ED}}$ (the position of TeV brane). By assuming all the Standard Model particles to be bounded on the TeV brane, they found that the mass of the Higgs boson could be expressed as~\cite{Randall1}
\begin{eqnarray}
m_{H}=e^{-k\pi R_{\text{ED}}}m_{\ast},
\end{eqnarray}
where $m_{\ast}$ is the bulk mass of the Higgs boson. They soon realized that it is not necessary to impose the bulk Planck mass to be the electroweak scale. On the contrary, the relaxed constraints on $k$ and $R_{\text{ED}}$, i.e., $k\sim M_{\ast}$ and $R_{\text{ED}}\sim10/M_{\ast}$, could efficiently suppress the bulk mass of the Higgs boson from the Planck scale on the Planck brane to the electroweak scale on the TeV brane while keeping the bulk Planck mass at the same scale as the Planck mass all the time. In this case, the hierarchy problem is solved well.

Indeed, the exponential warp factor is a crucial feature in extra-dimensional theories. As we have mentioned before, there is a serious problem left in domain wall model, which could be simply boiled down to a contradiction between the spectrum of KK gravitons and four-dimensional gravity on the domain wall~\cite{Randall2}. The contradiction seems to forbid people to construct an extra-dimensional theory with infinite extra dimensions. Soon after publishing RS-I model, Lisa Randall and Raman Sundrum realized that the warp factor introduced in RS-I model might be the key to this problem. In their famous RS-II model~\cite{Randall2}, they set the TeV brane to infinity and assumed that the Standard Model particles are bounded on the Planck brane. Then the spectrum of the unbounded KK gravitons becomes continuous and the Newtonian potential is modified as follows~\cite{Randall2}
\begin{eqnarray}\label{potential1}
V(r)&\sim&\frac{m_{1}m_{2}}{r}+
\int^{\infty}_{0}dm\frac{m}{k^{2}}\frac{m_{1}m_{2}e^{-mr}}{r}\nonumber\\
&\sim&\frac{m_{1}m_{2}}{r}\Big(1+\frac{1}{r^{2}k^{2}}\Big),
\end{eqnarray}
where $m_{1}$ and $m_{2}$ are the masses of two particles, $r$ is the distance between them, and the last term is contributed from the continuous KK modes. Note that the contribution from the massive KK gravitons will become significant if the distance $r$ is smaller than the Planck length. The Newtonian potential could be recovered when $r$ is relatively large. Therefore, a higher-dimensional theory with infinite extra dimensions could also obtain a four-dimensional effective theory by introducing a proper warp factor.

\section{Sources and spectra of gravitational waves}
\label{sec2}

It is known that GWs could be produced by any object with mass and acceleration. According to the characteristics of GWs, one can divide them into four categories: continuous GWs, compact binary GWs, stochastic GWs, and burst GWs. Continuous GWs are usually produced by massive objects with a spin, and their prominent feature is the long-lasting and constant frequency. Burst GWs refer to unknown or unanticipated GWs with a short duration, which represent new physics or unknown matters. Since we do not have any relevant observation data yet, the research on these two kinds of GWs has not received much attention. As for compact binary GWs, they are usually instantaneous and strong. Due to this feature, it is ``easy" to extract this kind of GW signal from noise signals. Therefore, compact binary GWs have always been valued in both theoretical and experimental fields. Now people have already detected several compact binary GWs and accumulated precious data. At last, we know that stochastic GWs involve the primordial universe. The detection of stochastic GWs is of great significance to our understanding of the evolution of the early universe, so stochastic GWs are also a subject worthy to study.

In extra-dimensional theories, most sources of GWs are similar to the case of four-dimensional space-time. But the corresponding GW spectra need to be corrected due to the existence of extra dimensions. In this section, we introduce the following sources and spectra of GWs in extra-dimensional theories: primordial universe, phase transitions, cosmic (super-)strings, and binary systems. These GWs have been extensively studied in four-dimensional space-time, so we just briefly present their different features in extra-dimensional theories.

For the case of the primordial universe, most research focuses on the GWs produced during the inflation occurring on our three-dimensional brane~\cite{Sahni:2001qp,Gorbunov:2001ge,Easther:2003re,Im:2017eju}. In order to discriminate higher-dimensional GWs from four-dimensional GWs, we need to study the evolution of GWs~\cite{Koyama:2000cc,Hiramatsu:2003iz,Easther:2003re,Hiramatsu:2004aa,
Seahra:2006tm,Guzzetti:2016mkm}. In Ref.~\cite{Langlois:2000ns}, the authors considered a five-dimensional anti-de Sitter space-time and the GWs formed during the slow-roll inflation. They found that at high energy (during the inflation) GWs could extend into the bulk and the amplitude of GWs on the brane is enhanced, which is different from the usual four-dimensional result. But at low energy, the spectra of GWs will be recovered to the case of the standard GR, which means that, in the current cosmic environment, it is difficult to detect extra dimensions with GWs for this extra-dimensional model. Some similar studies of higher-dimensional GWs in the primordial universe can also be found in Refs.~\cite{Giudice:2002vh,Frolov:2002qm,Kobayashi:2003cn,Hiramatsu:2003iz,
Sami:2004xk,BouhmadiLopez:2004ax,Ghayour:2012nf}. The Gauss-Bonnet effect on the spectra of higher-dimensional GWs was discussed in Refs.~\cite{Dufaux:2004qs,Bouhmadi-Lopez:2013gqa}. And some related numerical calculations can be found in Refs.~\cite{Ichiki:2004sx,Hiramatsu:2004aa,Kobayashi:2005jx,
Kobayashi:2005dd,Hiramatsu:2006bd}.

Although for different extra-dimensional models, the corrections to the GW spectra formed in the primordial universe are different, these modified spectra generally have two properties in common: during the primordial universe the effect of extra dimensions on GW spectra would be amplified because of the high energy (for other modified gravity theories, the corresponding GW spectra are generally independent of energy); at low energy (i.e., these GWs have evolved into the late universe) they are mostly identical to the standard four-dimensional result;

For the inflation caused by the dynamics of the inflaton in the bulk, there is hardly much research on GWs. The difficulties lie in two aspects: the dynamics of the bulk inflaton and the analysis of the perturbation~\cite{Koyama:2003yz}.

In the case of first-order phase transitions (three major sources of first-order phase transitions: collision of bubbles, turbulence in the primordial plasma, and magnetic field) in extra-dimensional theories, there could exist a strong signature in GWs~\cite{Randall:2006py,Hogan:2006va,Konstandin:2010cd,Chen:2017cyc,
Megias:2018sxv}, which provides a very useful means of detecting extra dimensions. For RS-I model, one can find a particular type of relic stochastic GW occurring at a temperature in the TeV range through a cosmological phase transition from an AdS-Schwarschild phase to the RS-I phase~\cite{Randall:2006py}. If the phase transition is strong enough, it is promising for the LISA detector to detect such a strong GW signal. In Ref.~\cite{Megias:2018sxv}, the authors considered a five-dimensional warped model including a scalar potential. They found that the stochastic GWs generated by phase transitions can be observed both at the LISA and the Einstein Telescope.

In most extra-dimensional models, there are new sources of first-order phase transitions. The GWs from phase transitions are promising to be detected in the regime where the parameters are justified, which is consistent with many four-dimensional theories. The problem is how to distinguish these signals. In extra-dimensional models, the characteristics of the phase transition GWs, compared with the phase-transition GWs in the standard GR, may be remarkably obvious (by adjusting the parameters in models). But if we compare these GWs with the GWs given by other modified gravity theories, we will find that the properties of these extra-dimensional GWs can be replaced by the GWs in other modified gravity theories. Therefore, at present, the GWs from phase transitions are not very suitable for detecting extra dimensions. For more related research, one can refer to Refs.~\cite{Hogan:2006va,Konstandin:2010cd,Chen:2017cyc} and references therein.

In order to distinguish higher-dimensional GWs from the GWs in the standard GR, some authors also studied the GWs generated by cosmic (super-)strings in extra-dimensional theories~\cite{OCallaghan:2010rlo,OCallaghan:2010mtk,OCallaghan:2010jrm}. There are two special bursts of gravitational radiation from cosmic (super-)strings. They are produced by the extreme kinematic events in the loop motion, known as cusps and kinks~\cite{Damour:2001bk,Damour:2004kw}. It was found that the impact of extra dimensions (which could be regarded as additional dynamical degrees of freedom) on the GW signals from cusp events is remarkable. The extra dimensions make the cusps more rounded and reduce the possibility of their formation. Therefore, due to the extra dimensions there exists a potentially significant damping on the GW signals from cusps~\cite{OCallaghan:2010rlo,OCallaghan:2010mtk}. With the improvement of the detection accuracy in the future, we are promising to detect the GWs from cusp events, which will provide effective constraints on extra-dimensional models. For the GW signals from kinks on cosmic (super-)strings, they are also suppressed in extra-dimensional theories. But, the suppression is not as significant as the case of cusps. Therefore, it could not provide a better chance of detecting extra dimensions (since the incidence of kinks on (super-)strings is relatively high, it is usually used for detecting cosmic (super-)strings~\cite{OCallaghan:2010jrm}). Similar to other stochastic GWs, the GWs from cosmic (super-)strings are also extremely weak (compared with detector noise) and it is difficult to distinguish the stochastic GWs in extra-dimensional models from the corresponding GWs in four-dimensional theories. There is still a long way to detect extra dimensions with stochastic GWs.

In fact, the studies on GWs from binary systems are the most extensive, especially after the first detection of GWs. These studies mainly focus on two aspects: applications in cosmology and astrophysics, and constrains on modified gravity theories. For extra-dimensional theories, we have mentioned that the correction to GWs can be reflected in the attenuation of amplitude. In general, the number and size of extra dimensions are the main factors, and then the configuration of extra dimensions. For almost all extra-dimensional models, the attenuation of the amplitude of GWs is faster than the case of four-dimensional theories. Therefore, the measurement of the amplitude attenuation of GWs can impose constraints on almost all extra-dimensional models. For stochastic GW signals, since they are very weak and it is difficult to locate their sources, it has little hope of studying how their amplitude decays. However, the GW signals generated by binary systems are ``strong'' and it is ``easy" to find the locations of their sources.

In addition, since the GWs generated by binary systems are usually accompanied by electromagnetic signals or neutrino signals, one can get more useful information by comparing these signals simultaneously. For extra-dimension theories, we can compare the order of the received signals to determine the size of extra dimensions or the values of other parameters. The studies on the GWs from binary systems will be presented in detail later.

The GWs mentioned above are the generalizations of four-dimensional GWs in extra-dimensional theories. For almost all modified gravity theories, one can study the properties and applications of these GWs under the corresponding gravity theory. However, in extra-dimensional models, besides these GWs there are a small number of GWs radiated by special sources. Here we introduce one of them: black strings.

A black string is regarded as a brane-world black hole in the bulk, which is a line singularity connecting our brane and a shadow brane in the bulk~\cite{Hirayama:2001bi,Kanti:2002fx,Kanno:2003au,Kudoh:2006bp,
Gergely:2006ei}. With regard to black strings, the most promising events involving GWs are the mergers of black strings and the perturbations of black strings (for example, a black string perturbed by an orbiting point like object). This type of GW is usually characterized by a discrete and high-frequency (about $\geq$ 300GHz~\cite{Li:2003tv,Li:2008qr,Arvanitaki:2012cn,Jr.:2005qbr}) spectrum~\cite{Seahra:2004fg,Clarkson:2006pq,Seahra:2009gw}. The discrete spectrum is formed because of the discrete tower of massive KK modes. The frequency of the spectrum is determined by the bulk curvature radius $\mathscr{L}$ and the brane separation distance $\mathscr{D}$. Generally speaking, the smaller the value of $\mathscr{D}/\mathscr{L}$, the higher the frequency of the corresponding GW is. In most of the extra-dimensional models, the GWs emitted by black strings have high-frequency spectra~\cite{Seahra:2004fg,Clarkson:2006pq,Seahra:2009gw}. These discrete high-frequency GWs could be a very important tool for probing extra dimensions. As far as we know, at present, there is no four-dimensional gravity theory which could produce such a GW spectrum, especially a discrete one. If a discrete GW spectrum is detected at the future high-frequency GW detectors, it would be a strong evidence of the existence of extra dimensions.

\section{shortcuts in extra-dimensional theories}
\label{sec3}

In extra-dimensional theories, it is generally accepted that GWs can propagate throughout the higher-dimensional space, while the other substances (our observable universe) are trapped on a three-brane. The speed of GWs in different extra-dimensional models may have different values, and there are many factors that can influence the propagation of gravity in the bulk, such as the size and number of extra dimensions. But most studies (which mainly refer to the research explaining some cosmological problems with extra-dimensional theories) suppose in advance that the speed at which GWs travel in the bulk is equal to the speed of light on the brane. So, in these studies, the trajectories of gravitons are the null geodesics in the bulk. In this section, we will introduce some of these studies, especially those related to the shortcuts of GWs.

We first introduce geodesics and ``fifth force'' in extra-dimensional theories. For black holes in RS models~\cite{Randall1,Randall2}, Andrew Chamblin {\it et al.} have investigated time-like geodesics and null geodesics in Ref.~\cite{Chamblin:1999by} based on the Schwarzschild-anti-de Sitter solution, which offers valuable guidance on calculating geodesics in extra-dimensional models with large extra dimensions (see also Refs.~\cite{Chamblin:2000md,Zhou:2011iq}). In the early days, the null geodesics in extra-dimensional theories were studied in order to solve the horizon problem in a different way than inflation~\cite{Chung:1999xg,Chung:1999zs,Ishihara:2000nf}. In these studies, the role of extra dimensions was ignored when considering the motion of the matter on the brane. It was later discovered that, in general brane background, the geodesics of the massive particles on the brane are also affected due to the presence of extra dimensions~\cite{Mueck:2000bb}. Such effect manifests as an extra non-gravitational force acting on the massive particles on the brane (see Refs.~\cite{Youm:2000ax,Youm:2001qc,Magpantay:2011jy} and references therein). In some literature, this new dynamical force due to extra dimensions is also directly called the ``fifth force'' (see Refs.~\cite{Gross:1983hb,Kovacs:1984qx,Gegenberg:1984gx,Mashhoon:1998tp,
Wesson:1999mc}). In Ref.~\cite{Magpantay:2011jy}, the author found that the fifth force does not change the velocity of the particles on the three-brane but their masses, while for the particles in the bulk, their motions would result in a time-dependent Heisenberg uncertainty principle.

According to the development of our current experiments, if this force really exists, it is generally negligible compared to the other four known forces. Therefore, when we consider most problems in extra-dimensional models, we think that the geodesics of the particles on the brane are the same as the result of the standard GR, and we do not need to calculate them with the induced metric on the brane. But for the gravity traveling in the bulk, since it is very weak and the measurement about gravity, especially GWs, is not accurate enough, the influence of extra dimensions on the propagation of gravity cannot be easily ignored.

For most extra-dimensional theories, the null geodesics in the bulk are usually not the same as the null geodesics on the three-brane, so the observer on the brane could perceive that the propagations of gravitational and electromagnetic signals on the brane have differences in velocity and amplitude attenuation. After the GW150914 event, some people started to study how to limit extra-dimensional models by comparing the null geodesics of GWs in the bulk with the null geodesics of light on the brane, and the core issue of the research is the shortcuts of GWs.

Earlier we mentioned that extra-dimensional theories can be used to solve the horizon problem by the shortcut of gravity. The so-called ``shortcut'' results from the fact that since the null geodesics of GWs in the bulk are different from the null geodesics of light on the brane, observers on the brane may get an illusion that gravity is faster than light or gravity is a ``superluminal'' interaction.
Since there appears a ``superluminal'' interaction on the brane, then we naturally have to ask if the ``superluminal'' interaction would lead to a violation of causality? In most cases, it is possible for the observers on the brane to observe an apparent causality violation~\cite{Ishihara:2000nf,Stoica:2001qe,Abdalla:2002ki,Pas:2006si}. However, in a five-dimensional space-time (similar to the KK theory), if the factorizable ansatz for the bulk metric satisfies the requirements that all components in the metric are independent of the fifth coordinate and the component $G_{55}$ is a constant, then we can avoid the apparent violation of causality~\cite{Parthasarathy:2012jx}. In addition, one can also use the causality to constraint the GWs in the bulk. In RS-II model, according to the causal structure of the flat brane universe, one can obtain some boundary conditions for the GWs in the bulk~\cite{Ichiki:2003hf}. Other related studies can be found in Refs.~\cite{Chung:1999zs,Kaelbermann:1999jw,Chung:1999xg,Abdalla:2005wr}.

Now let us take a look at how to calculate a shortcut of gravity in a specific extra-dimensional model. Here we mainly present a kind of scenario, which was proposed by Robert R. Caldwell and David Langlois~\cite{Caldwell:2001ja}. The background of the model is a Schwarzschild-anti-de Sitter space-time. In this model, gravitons can propagate in the infinite and warped (bulk) spece-time, but photons are confined on a three-brane. The bulk metric is given by ~\cite{Birmingham:1998nr,Mann:1996gj,Brill:1997mf,Vanzo:1997gw,Caldwell:2001ja}
\begin{eqnarray}\label{Metric1}
ds^2_{\texttt{bulk}1}=-f(R)dT^2+f(R)^{-1}dR^2+R^2d\Sigma_{k}^{2}.
\end{eqnarray}
Here $d\Sigma_{k}^{2}$ is a metric on a three-dimensional surface of constant curvature $k$ (be careful not to confuse the new symbols with the ones used earlier). The expression of $f(R)$ is assumed to be
\begin{eqnarray}\label{fR}
f(R)=k+\frac{R^2}{l^2}-\frac{\mu}{R^2},
\end{eqnarray}
where $l$ ($>0$) is the constant curvature radius (which can be considered as the size of the extra dimension) and $\mu$ is the Schwarzschild-like mass.

\begin{figure}[htb]
\begin{center}
\includegraphics[width=12cm]{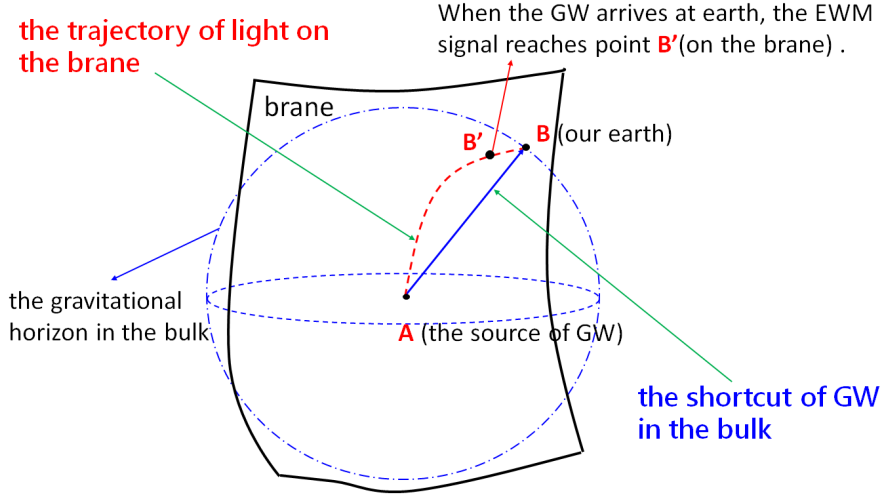}
\end{center}
\caption{\label{shiyitu} The trajectories of GW and electromagnetic wave (EMW)~\cite{Yu:2016tar}. The points $A$, $B$, and $B'$ are all on the brane. The dashed red line $AB'B$ represents the track of the null geodesic on the brane and the solid blue line AB is the track of the null geodesic in the bulk.}
\end{figure}

The induced metric on the brane is given as
\begin{eqnarray}\label{Metric2}
ds^2_{\texttt{brane}}=-dt^2+R_b(t)^2d\Sigma_{k}^{2},
\end{eqnarray}
with which one can calculate the horizon radius for the propagation of light on the brane. Considering any two points $A$ and $B$ on the brane (both $A$ and $B$ represent spatial points), they can be connected either by a null geodesic on the brane or a null geodesic in the bulk and usually the two null geodesics are different (see Fig.~\ref{shiyitu} quoted from Ref.~\cite{Yu:2016tar}). Using a spherical coordinate system in the brane and setting the coordinate origin at the point $A$ (the corresponding time is marked as $T_A$), one can ignore the angular variables naturally. The geodesics in the bulk can be described by a three-dimensional metric:
\begin{eqnarray}\label{Metric3}
ds^2_{\texttt{bulk}2}=-f(R)dT^2+f(R)^{-1}dR^2+R^2dr^{2},
\end{eqnarray}
where $r$ is the radial coordinate. Utilizing the Killing vectors of the metric and the nature of null geodesics, the comoving distance from point $A$ (time $T_A$) to point $B$ (time $T_B$) can be obtained. In the case of $k=\mu=0$, the result is simplified to
\begin{eqnarray}\label{gw1}
r_{gAB}=\left(\left[\int^{T_B}_{T_A}\frac{dt}{a}\sqrt{1+l^2H^2}\right]^2-
\left[\int^{T_B}_{T_A}\frac{dt}{a}lH\right]^2\right)^{1/2},
\end{eqnarray}
where $a$ and $H$ are the scale factor and Hubble constant on the brane.

For the light on the brane traveling from time $T_A$ (also the position $A$) to time $T_B$, the position it arrives is not necessarily the point $B$. One can temporarily suppose it is point $B'$. With the induced metric, the comoving distance of light is
\begin{eqnarray}\label{photon1}
r_{\gamma AB'}=\int_{T_A}^{T_B}\frac{dt}{a}.
\end{eqnarray}

Then comparing $r_{gAB}$ and $r_{\gamma AB'}$, we can determine whether the apparent ``speed'' of the GWs on the brane is superluminal. According to the results of Ref.~\cite{Caldwell:2001ja}, it is not easy to solve the horizon problem in this scheme because gravity is not much ``faster'' than light. There are some other works trying to solve the horizon problem in extra-dimensional models, but the conclusions are similar~\cite{Chung:1999xg,Chung:1999zs,Abdalla:2002ki,Abdalla:2002je,
Abdalla:2005wr}.

It is worth mentioning that not all GWs in extra-dimensional theories can take shortcuts. In certain extra-dimensional models (such as RS models), there is no shortest path in the bulk and the shortest cut is only present on the brane~\cite{Abdalla:2002ir}. In Ref.~\cite{Abdalla:2001he} the authors also pointed out that the existence of shortcuts depends on a set of conditions in a six-dimensional brane-world model.

The analysis and calculation above are based on two points: the null geodesics of GWs in the bulk and the null geodesics of light on the brane are different, and the speeds of GWs and light are the same constant. Are these two points appropriate for all extra-dimensional models? The former, even in the absence of calculation, is basically accepted by all researchers. But for the latter, it cannot be taken for granted.

The extra-dimensional models assuming that the speed of gravity equals the speed of light, usually have a common feature that they do not have an effective description of Lorentz invariant. If we consider gravitational Lorentz violation, the speed of GWs in the bulk does not have to be equal to the speed of light. For example, in an asymmetrically warped higher-dimensional space-time, the speed of light is not fixed due to the asymmetric warped extra dimensions, and the speed of gravity along the brane also varies over the distance on the brane. Only at a large scale, there will be a clear gap between their speeds. And the speed of gravity is always greater than the speed of light (see Ref.~\cite{Csaki:2000dm} for more details). Another example: for the model considered in Ref.~\cite{Ahmadi:2012we}, the authors found that only if the energy density of the matter localized on the brane vanishes, the maximum speed in the bulk (i.e., the speed of GWs) could be equal to the speed of light on the brane. In other cases, the maximum speed in the bulk is faster than the speed of light. Similar conclusions were also obtained in Refs.~\cite{Gogberashvili:2006dz,Ahmadi:2012we}.

\section{Size and number of extra dimensions}
\label{sec4}

We have already introduced the sources of GWs, the characteristics of the corresponding spectra, and the shortcuts of GWs in various extra-dimensional theories. In this section, we will focus on two important applications of GWs in extra-dimension theories, i.e., utilizing GWs to detect the size and number of extra dimensions.

\subsection{Size of extra dimensions}
\label{sec4.1}

The presence of extra dimensions has some minor corrections to the spectra of stochastic GWs, and the corrections are usually directly related to the size of extra dimensions, which provides a method to estimate the size of extra dimensions. However, for current observation precision, it is impossible to use stochastic GWs to detect extra dimensions, because they are too weak to be detected. In the future, with the development of experimental technology, if we can accurately extract the data of stochastic GWs from noise, then there is no doubt that these data are crucial information for detecting extra dimensions. Therefore, it is necessary to study theoretically the relation between the spectra of stochastic GWs and the size of extra dimensions. The spectra have different properties to different extra-dimensional models. We introduce a representative model here.

In Ref.~\cite{Giudice:2002vh}, the authors considered a five-dimensional brane-world model. They found that for the GWs generated during the inflation, there exists a correction term proportional to $(HR_{\text{ED}})^2$. Note that $H$ is the Hubble constant and $R_{\text{ED}}$ is the size of the extra dimension. If this model is the real model of the universe, the size of the extra dimension could be determined accurately by measuring the background spectrum of stochastic GWs. Of course, if the observation of stochastic GWs is consistent with the prediction of GR, then extra dimensions may not exist ($R_{\text{ED}}=0$). In order to study the effect of the number and structure of extra dimensions on stochastic GWs, the authors also generalized their discussion to a model with multiple extra dimensions and the warped RS models, respectively (see details in Ref.~\cite{Giudice:2002vh}). Similar works on studying the influence of extra dimensions on the spectra of stochastic GWs can be found in Refs.~\cite{Hogan:2000aa,McWilliams:2009ym,Yagi:2011yu}.

The premise of detecting the size of extra dimensions in this way includes two aspects: the stochastic GWs detected in the future deviate from the results of GR and the correction to the spectra of stochastic GWs is due to extra dimensions. But in fact, even if the stochastic GWs we observe is inconsistent with the prediction of GR, it is difficult to judge whether the correction is due to extra dimensions or other modified gravity theories. In order to distinguish higher-dimensional GWs from other four-dimensional GWs, we need to find out more unique properties of extra-dimensional models, such as the aforementioned discrete GW spectrum and the shortcut of GWs. Next, we introduce another way to detect the size of extra dimensions in light of GWs and their electromagnetic counterparts. This method could rule out the modified gravity theories which do not possess shortcut effect.

In Sec.~\ref{sec3}, we have mentioned that when there exist extra dimensions, the propagation paths of GWs and EMWs are different. And usually the observers on the brane will feel that GWs run faster than EMWs. For a long time, since we did not observe GWs directly, physicists did not have a consistent view about whether GWs are faster than EMWs for the observers on the brane. In addition, the relevance of the two signals is also somewhat controversial. As far as we know, theoretical research has begun at the end of the last century~\cite{Paczynski:1986px,Meegan:1992xg,Cutler:1992tc,Kochanek:1993mw,Finn:1999vh}.

In the GW150914 event, just 0.4 seconds after GW150914 was detected, the Fermi GBM captured a gamma ray burst. Since the sky location of its source is close to the source of GW150914, some people believe that it is an electromagnetic counterpart of GW150914. However, the others suspect it is a coincidence because the positioning range of GW150914 is too large and vague. Regardless of this disputation, if these two signals are generated at the same time by the merger of a pair of black holes, then it is necessary to explain why there exists a time delay between them. Considering the shortcuts of GWs in extra-dimensional theories, we naturally think of using this property of GWs to explain the phenomenon of the time delay. On the other hand, this event also provides an opportunity to constrain the parameters in extra-dimensional models.

Let us first look at a very intuitive example. Imagine our universe is a spherical shell. EMWs can only move on the shell, but GWs can travel through the shell. Given any two points on the shell, there is a shortest line in the bulk connecting them, which is the trajectory of GWs. Similarly, we can also find a shortest route on the shell to connect them, which is the trajectory of EMWs. Intuitively, it is clear that the former is shorter than the latter. Since EMWs and GWs have the same speed, observers on the shell will find that GWs travel faster than EMWs. Combining this physical image with the GW150914 event, the size of the spherical brane-universe is estimated at about $10^{30}$km~\cite{Gogberashvili:2016kgj}.

Although the shell-universe is physically intuitive, this model is too rudimentary. A more general bulk space-time is Schwarzschild-anti-de Sitter space-time. The metric is also given by Eq.~(\ref{Metric1}). Therefore, from the time $T_A$ to the time $T_B$, the comoving distance traveled by GWs (for the simplest case $k=\mu=0$) should be given by Eq.~(\ref{gw1})~\cite{Caldwell:2001ja}. This comoving distance is also called gravitational horizon radius. Comparing this radius with the comoving distance of light on the brane (see Eq.~(\ref{photon1})), one can obtain the difference between these two distances within a given time interval. Based on the analysis above (see also in Ref.~\cite{Caldwell:2001ja}), Hao Yu and Yu-Xiao Liu {\it et al.} tried to restudy the model with the data of the GW150914 event~\cite{Yu:2016tar}. In order to study the effect of the curvature $k$ on the propagations of GWs and EMWs, the authors considered a more general case with $k\neq0$.

For the de Sitter model of the universe, the gravitational horizon radius of GWs is~\cite{Yu:2016tar}
\begin{eqnarray}\label{rg2}
r_{gAB}=\frac{1}{ \sqrt{k}}\Bigg[\arctan\Bigg(\frac{\sqrt{k}(1+z)}
{\sqrt{H_B^2-kz(2+z)}}\Bigg)
-\arctan\Bigg(\frac{\sqrt{k}}{H_B}\Bigg)\Bigg],
\end{eqnarray}
where $H_B$ is the value of the Hubble parameter at time $t_B$ and $z$ is the redshift satisfying $1+z=\frac{a(t_B)}{a(t_A)}$. It can be seen that $r_{gAB}$ has no concern with the constant curvature radius $l$. The comoving distance of light is given by
\begin{eqnarray}\label{rlAB}
r_{\gamma AB'}
=\frac{1}{\sqrt{k}}\sin\Bigg[\arctan\Bigg(\frac{\sqrt{k}}
{\sqrt{-k+(1+z)^{-2}(H_B^2+k)}}\Bigg)
-\arctan\Bigg(\frac{\sqrt{k}}{H_B}\Bigg)\Bigg],
\end{eqnarray}
which is also not a function of the parameter $l$. Therefore, the value of the parameter $l$ could not lead to any difference between the comoving distances of a GW signal and an EMW signal. When the curvature $k$ approaches to zero, $r_{gAB}=r_{\gamma AB'}=\frac{z}{H_B}$ (see details in Refs.~\cite{Ishihara:2000nf,Caldwell:2001ja,Yu:2016tar}).

The authors found that as the distance of the GW source increases ($z$ increases), the influence of the non-vanishing $k$ on the comoving distances of GWs and EMWs becomes more and more significant. For explaining the 0.4 second delay between the gravitational signal and the electromagnetic signal in the GW150914 event, one needs the value of $k$ to satisfy $k\sim10^{-50}$. It is a very small value, which completely reaches to the requirement of the current observation ($k$ should be smaller than $10^{-4}$).

In the Einstein-de Sitter model of the universe, the comoving distance of light on the brane is similar to Eq.~(\ref{rlAB}), but the gravitational horizon radius is much more complicated than Eq.~(\ref{rg2}) (see Eqs.~(20), (29), and (30) in Ref.~\cite{Yu:2016tar}). Although the size of the extra dimension still does not affect the propagation of the electromagnetic signal on the brane (see Eq.~(30) in Ref.~\cite{Yu:2016tar}), it has impact on the gravitational horizon radius in the bulk (see Eqs.~(20) and (29) in Ref.~\cite{Yu:2016tar}). However, contrary to expectations, substituting the data of the GW150914 event into the formulas, it can be found that the radius of the extra dimension has little impact on the propagation of GWs. Therefore, it is almost impossible to determine the size of the extra dimension with the rough data of the GW150914 event. Since the size of the extra dimension has little effect on the propagation of GWs, the discussion and conclusion about $k$ are the same as those in the de Sitter model of the universe.

If the electromagnetic signal measured in the GW150914 event was questioned, then the emergence of the GW170817/GRB170817A event might dispel many people's doubts. In the light of the GW170817/GRB170817A event, the authors in Ref.~\cite{Visinelli:2017bny} considered the same Schwarzschild-anti-de Sitter space-time with the metric (\ref{Metric1}). Performing the time lag between GW170817 and its counterpart GRB170817A, they determined an upper bound on $l~(\leq0.535\,\text{Mpc})$ at $68\%$ confidence level.

\subsection{Number of extra dimensions}
\label{sec4.2}

To our best knowledge, there are not many studies on the application of GWs to detect or limit the number of extra dimensions. The following properties of GWs are directly related to the number of extra dimensions: anomalous polarization amplitude~\cite{Deffayet:2001uk,Alesci:2005rb},\footnote{In many modified gravity theories, there exist extra polarizations comparing with the two transverse quadrupolar $(+~\times)$ modes of GR~\cite{Will:2005va}. For GWs in extra-dimensional theories, anomalous polarizations also exist because the radiation sources of KK gravitons are various.} leakage of GWs into extra dimensions~\cite{Gregory:2000jc,Deffayet:2007kf,Pardo:2018ipy,Abbott:2018lct}, and other corrections to GWs~\cite{Hogan:2000aa,Qiang:2017luz}. In this section, we only pay attention to the leakage of GWs.

Before GWs were detected, the most common experiment (except high-energy experiments) to detect extra dimensions was to measure the relation between gravitational potential and the distance between test particles. For ADD model~\cite{ArkaniHamed1}, when the scale (the distance between test particles) is much smaller than the size of extra dimensions, according to Gauss's law in $D=(4 + d)$ dimensions, the gravitational potential can be written as
\begin{eqnarray}
V(r)\sim\frac{m_1m_2}{M^{d+2}_{\text{Pl}(4+d)}}\frac{1}{r^{d+1}},
\end{eqnarray}
where $r$ is the distance between two test particles of mass $m_1$, $m_2$. And if $r$ is much larger than the size of extra dimensions, one can get the usual $1/r$ gravitational potential:
\begin{eqnarray}
V(r)\sim\frac{m_1m_2}{M^{d+2}_{\text{Pl}(4+d)}R_{\text{ED}}^d}\frac{1}{r}.
\end{eqnarray}
The most accurate gravitational potential experiment currently measured in the laboratory indicates that the gravitational potential between two objects at the submillimeter range, still satisfies the Newtonian law ($V(r)$ is proportional to $1/r$)~\cite{YangLuo:2012PRL,TanLuo:2016PRL}. Therefore, in order to restore the effective four-dimensional gravitational potential (when $r> 0.1\,\text{mm}$) in ADD model, we require that the scale of extra dimensions is less than $0.1\,\text{mm}$ and the number of extra dimensions must also be consistent with the experimental data~\cite{ArkaniHamed1}. Besides ADD model, in most extra-dimensional models, the gravitational potential on the brane will also be corrected when the distance between test particles is below the scale of extra dimensions~\cite{Randall1,Barvinsky:2003jf}. The correction usually results in the phenomenon that the gravitational potential on the brane is weaker than the result of GR.\footnote{Of course, there are exceptions. For example, in Ref.~\cite{Kogan:2000cv}, the authors studied an extra-dimensional model interpolating between Bi-gravity model~\cite{Kogan:1999wc} and GRS model~\cite{Gregory:2000jc}, they found that the gravity on the brane is not the effective four-dimensional gravity at small and very large scales (about $10^{26}\,\text{cm}$).} Such an effect on the gravitational potential also applies to GWs in extra-dimensional models, which is phenomenologically described as the leakage of GWs into extra dimensions. Next, we discuss the leakage of GWs in a kind of extra-dimensional model with a new length scale.

In some extra-dimensional models (such as DGP gravity, a brane-world model with an infinite extra dimension~\cite{Dvali:2000hr}), there exists a screening scale $R_c$. The reason why it is called screening scale is that beyond this scale gravity will deviate from GR obviously. At the scale below $R_c$ (such as in the solar system), gravity must pass the standard tests of GR. For the extra-dimensional model researched in Ref.~\cite{Deffayet:2007kf}, if the distance traveled by GWs is much larger than the screening scale $R_c$, the GW amplitude scale can be given as
\begin{eqnarray}\label{GWAM}
h_{+,\times}\propto R_{\text{OS}}^{-(D-2)/2},
\end{eqnarray}
where $R_{\text{OS}}$ is the distance between the observer on the brane and the source of GWs, and $D$ is the dimensionality of the bulk space-time. Therefore, the usual four-dimensional (i.e., $D=4$) GW amplitude scale is the standard $h_{+,\times}\propto R_{\text{OS}}^{-1}$. But if the distance that GWs propagate is shorter than the screening scale, then Eq.~(\ref{GWAM}) is no longer applicable. More generally, the GWs damping
with luminosity distance can be expressed as~\cite{Deffayet:2007kf,Pardo:2018ipy}
\begin{eqnarray}\label{GWAM1}
h_{+,\times}\propto \frac{1}{d_L[1+(\frac{d_L}{R_c})^{\tilde{n}(D-4)/2}]^{1/\tilde{n}}},
\end{eqnarray}
where $\tilde{n}$ determines the transition steepness and $d_L$ is the luminosity distance of GW source. When $d_L\gg R_c$, Eq.~(\ref{GWAM1}) reduces to Eq.~(\ref{GWAM}). It is foreseeable that the number of extra dimensions is an important parameter in Eq.~(\ref{GWAM1}) even $d_L< R_c$, which is the reason why we can use GWs to detect the number of extra dimensions in this model.

Now, combining with a GW event we present a concrete result. Based on the work of Deffayet and Menou~\cite{Deffayet:2007kf}, Kris Pardo {\it et al.} applied the leakage phenomenon of GWs to the GW170817 event~\cite{Pardo:2018ipy}. In their work, two theories have been studied with GW170817: an extra-dimensional theory with a screening scale and a theory with decaying gravitons. For the second case, the GW amplitude scale is given as
\begin{eqnarray}\label{GWAM2}
h_{+,\times}\propto\frac{\exp(-d_L/R_g)}{d_L}.
\end{eqnarray}
The parameter $R_g$ is the distance traveled by a graviton during the average time of decay~\cite{Pardo:2018ipy}). Here, we are only concerned about the first theory.

In the GW170817/GRB170817A event, there exists a time interval between the GW signal and its electromagnetic counterpart. One can use the time interval of the two signals to calculate $R_c$, which is a function of the dimensionality of the bulk space-time:
\begin{eqnarray}\label{GWAM3}
R_c=\frac{d_L^{\text{EMW}}}{[(\frac{d_L^{\text{GW}}}{d_L^{\text{EMW}}})
^{\tilde n}-1]^{\frac{2}{\tilde n(D-4)}}}.
\end{eqnarray}
Here $d_L^{\text{EMW}}$ is the luminosity distance derived from the EMW observation, which is assumed as the true luminosity distance: $d_L^{\text{EMW}}=d_L$. The GW luminosity distance is labeled as $d_L^{\text{GW}}(\neq d_L^{\text{EMW}})$. Then taking the data of the GW170817/GRB170817A event into Eqs.~(\ref{GWAM3}) and (\ref{GWAM1}) one can obtain the range of the value of parameter $D$. For the SHoES value of $H_0$ and the Planck value of $H_0$, the results are $D=4.02^{+0.07}_{-0.10}$ and $D=3.98^{+0.07}_{-0.09}$, respectively. Both of these results indicate that the GW170817/GRB170817A event does not support this extra-dimensional theory ($D\geq5$) (see more details in Ref.~\cite{Pardo:2018ipy})).

There are also some other corrections to GWs based on the number of extra dimensions. In Ref.~\cite{Hogan:2000aa}, the author studied the dynamical history and stabilization of one to seven extra dimensions. Since the spectra of GWs have different properties for different number of extra dimensions, one can combine these characteristics with GW data to determine the number of extra dimensions. In addition, the inspiral GWs from black hole binaries also have some properties which are closely related to extra dimensions~\cite{Qiang:2017luz}. For example, in a general KK theory, one can use perturbation analysis to get a first-order correction to the inspiral GWs formed by black hole binaries. Such a correction is due to the volume change of extra dimensions near the region of black hole binaries. The correction depends on a new parameter $\chi=\frac{d}{2+d}$, where $d$ is the number of extra dimensions. As an aside, the propagation velocity of GWs can also reflect the number of extra dimensions under certain conditions. Some related research can be found in Refs.~\cite{Cardoso:2002pa,Barvinsky:2003jf}.

\section{Other gravitational waves}
\label{sec5}

In this section, we introduce some other research related to GWs in extra-dimensional theories. If GWs pass through two stationary objects, their relative distance will exhibit a change, and the displacement may be permanent. This effect is called gravitational memory, which is first studied by Yakov B. Zel'dovich {\it et al.} in linearized gravity theories~\cite{Zeldovich:1974,Braginsky:1985}. In higher-dimensional space-time, this effect has also been investigated in Refs.~\cite{Chu:2016ngc,Hollands:2016oma,Garfinkle:2017fre,Mao:2017wvx,
Pate:2017fgt}.

In GR, the tidal Love number for black holes is zero and it could be nonzero in modified gravity theories, which is another window exploring extra-dimensional theories. The effect on the tidal Love number due to the presence of extra dimensions is given in Refs.~\cite{Chakravarti:2018vlt,Chakravarti:2019aup}. It is found that with multi-messenger observations of GWs, one can constrain the brane tension in some brane-world models~\cite{Chakravarti:2019aup}.

The ringing modes of black holes are also called quasinormal modes, which are usually used to describe how an asymmetry black hole evolves towards a perfect sphere. This process contains a lot of important information about black holes, and the information can spread out in the form of GWs. The effect of extra dimensions on the ringing modes of black holes resides in the gravitational perturbation equation by introducing massive perturbation modes. If the massive gravitational perturbation modes can be observed, it would be a definitive evidence of the existence of extra dimensions. Compared to the massless mode in GR (the imaginary parts of the quasinormal-mode frequency of the massless mode are very small), the massive gravitational perturbation decays more slowly, which provides a new method for the detection of extra dimensions with GWs~\cite{Toshmatov:2016bsb,Chakraborty:2017qve,Aneesh:2018hlp}.

For the radiated power by GWs from a binary system, there exist corrections to the stellar period due to extra dimensions. In high energy regime, the author in Ref.~\cite{Garcia-Aspeitia:2013jea} got a correction term in the equation of period, which could be used to calculate a lower energy bound for brane tension.

Finally, most of the extra-dimensional models we introduced previously have only one extra dimension, but obviously GWs would possess different properties when there are multiple extra dimensions. We will not introduce them one by one here and readers can refer to the references mentioned earlier about multiple extra dimensions.

\section{Summary and outlook}
\label{sec6}

Up to now, more than ten GW events have been detected by the LIGO and Virgo Scientific Collaborations. These GW events open a new era in astronomy, cosmology and other physics~\cite{Blair:2016idv,Cai:2017cbj}. As a new powerful tool, GWs can also reveal the secret of extra dimensions.

In this review, we briefly described some features of GWs in various extra-dimensional models. We first introduced the development history of GWs and several important extra-dimensional theories. Then we showed some recent works focusing on the correction of GW spectrum in extra-dimensional theories. Several major GW sources and the corresponding spectra were discussed. Next, we reviewed the shortcut of GWs, which is one of the main characteristics of extra-dimensional theories. Using the shortcut and amplitude attenuation of higher-dimensional GWs, we discussed two important applications of GWs: constraining the size and number of extra dimensions. Finally, we listed some other studies about GWs in extra-dimensional theories.

For a long time, many people believed that the major breakthrough in the research on extra-dimensional theories relies on high-energy particle collisions. The construction of high-energy particle colliders is undoubtedly instructive for the study of extra-dimensional theories. But, the constraints from GW observations on extra-dimensional theories cannot be ignored. At present, our detection of GWs is still in infancy. We believe that the collection of more accurate data in the future will impose stricter restrictions on extra-dimensional models.

In addition to the research mentioned above, we need to continue going broader and deeper in this field. For example, we can concentrate on the study of burst GWs in extra-dimensional models (it is also a significant topic that many modified gravity theories should pay attention to). The GWs generated directly in the bulk would be a major source of burst GWs. In addition, as far as we know, there is almost no relevant literature studying the propagations of neutrinos, gravitons, and photons simultaneously in extra-dimensional models, which may also provide more information about extra dimensional models.

Due to the limitation of space, we are unable to discuss all aspects of GWs in extra-dimensional models. We tried our best to focus on the most-researched issues and list all the related literature we know. The references we quote do not represent all the research in this field, and some important literature may be omitted by us. We apologize for this.

\section*{Acknowledgments} \hspace{5mm}

This work was supported by the National Natural Science Foundation of China (Grants No. 11875151 and No. 11522541) and the Fundamental Research Funds for the Central Universities (Grants No. lzujbky-2018-k11). H. Yu was supported by the scholarship granted by the Chinese Scholarship Council (CSC).

\providecommand{\href}[2]{#2}\begingroup\raggedright

\end{document}